\documentstyle[12pt,epsf]{article}

\begin{document}


\begin{center}
\begin{verbatim}
                    OOOOO    OOOO     OOOO     OOOOO   
                    O        O   O    O   O    O       
                    OOOOO    O   O    O   O    OOOOO   
                    O        O   O    O   O    O       
                    OOOOO    OOOO     OOOO     OOOOO   
                       X        O        I       V     
                                        F       E      
                         C      U      F       N       
                                      R       T        
                           L    B    A       S         
                            U       C                  
                             S  L  T                   
                              I   I                    
                               VEV                     
                                                                     
                        Version 2.1 (2007)                   
\end{verbatim}

 {\Large \bf
\vskip 1cm
\mbox{EDDE Monte Carlo event generator.Version 2.1}
}
\vskip 2cm

\mbox{Petrov~V.A., Ryutin~R.A., Sobol~A.E.}

\mbox{{\small Institute for High Energy Physics}}

\mbox{{\small{\it 142 281} Protvino, Russia}}

\mbox{and}

\mbox{Guillaud~J.-P.}

\mbox{{\small Indiana University, USA}}

 \vskip 1.75cm
{\bf
\mbox{Abstract}}
  \vskip 0.3cm

\newlength{\qqq}
\settowidth{\qqq}{In the framework of the operator product  expansion, the quark mass dependence of}
\hfill
\noindent
\begin{minipage}{\qqq}

EDDE is a Monte Carlo event generator for different Exclusive and Semi-Inclusive Double Diffractive 
processes. The program is based on the extended Regge-eikonal approach for "soft" 
processes. Standard Model and its extensions are used for "hard" fusion processes.

\end{minipage}
\end{center}


\begin{center}
\vskip 0.5cm
{\bf
\mbox{Keywords}}
\vskip 0.3cm

\settowidth{\qqq}{In the framework of the operator product  expansion, the quark mass dependence of}
\hfill
\noindent
\begin{minipage}{\qqq}
Exclusive Double Diffractive Events -- Pomeron -- Regge-Eikonal model -- event generator
\end{minipage}

\end{center}

\setcounter{page}{1}
\newpage


\section{Introduction}

In the present paper we give a brife description of the Monte-Carlo event generator EDDE, which includes Exclusive Double Diffractive Events (EDDE, $p+p\to p+M+p$) and Semi-inclusive Double Diffractive Events (SI DDE, $p+p\to p+\{ X M Y\}+p$). Here M is the central particle or system of particles (Higgs boson, RS1 model particles, graviton, $\chi_{c,b}$, glueball, $Q\bar{Q}$, $gg$, $\gamma\gamma$, $Q\bar{Q}g$, $ggg$), X and Y denote "soft" gluon radiation in the central region, and "+" means Large Rapidity Gap.

 EDDE gives us unique experimental possibilities for particle searches
and investigations of diffraction itself. This is due to several advantages of the process: a) clear
signature of the process; b) possibility to use "missing mass method" that improve the mass 
resolution; c) background is strongly suppressed; d) spin-parity analysis of the central system can be 
done; e) interesting measurements concerning the interplay between "soft" and "hard" scales are possible~\cite{diffpatterns}. All these properties can be realized
in common CMS/TOTEM detector measurements at LHC~\cite{TDRS}.

SI DDE
is important as a source of main backgrounds for the
exclusive processes. 

Large number of event generators are devoted to partonic processes of the Standard Model and its extensions, i.e. work at small distances. It is wellknown, that perturbation theory has some problems in the description of processes at large distances. That is why diffractive processes are usually considered as special cases which description is based on different phenomenological approaches. The most popular approach is the Regge-eikonal model.


\section{Physics of EDDE}

\begin{figure}[ht]
\label{pp_pXp}
\mbox{a)}
\vbox to 5cm {\hbox to 7cm{\epsfxsize=7cm\epsfysize=5cm\epsffile{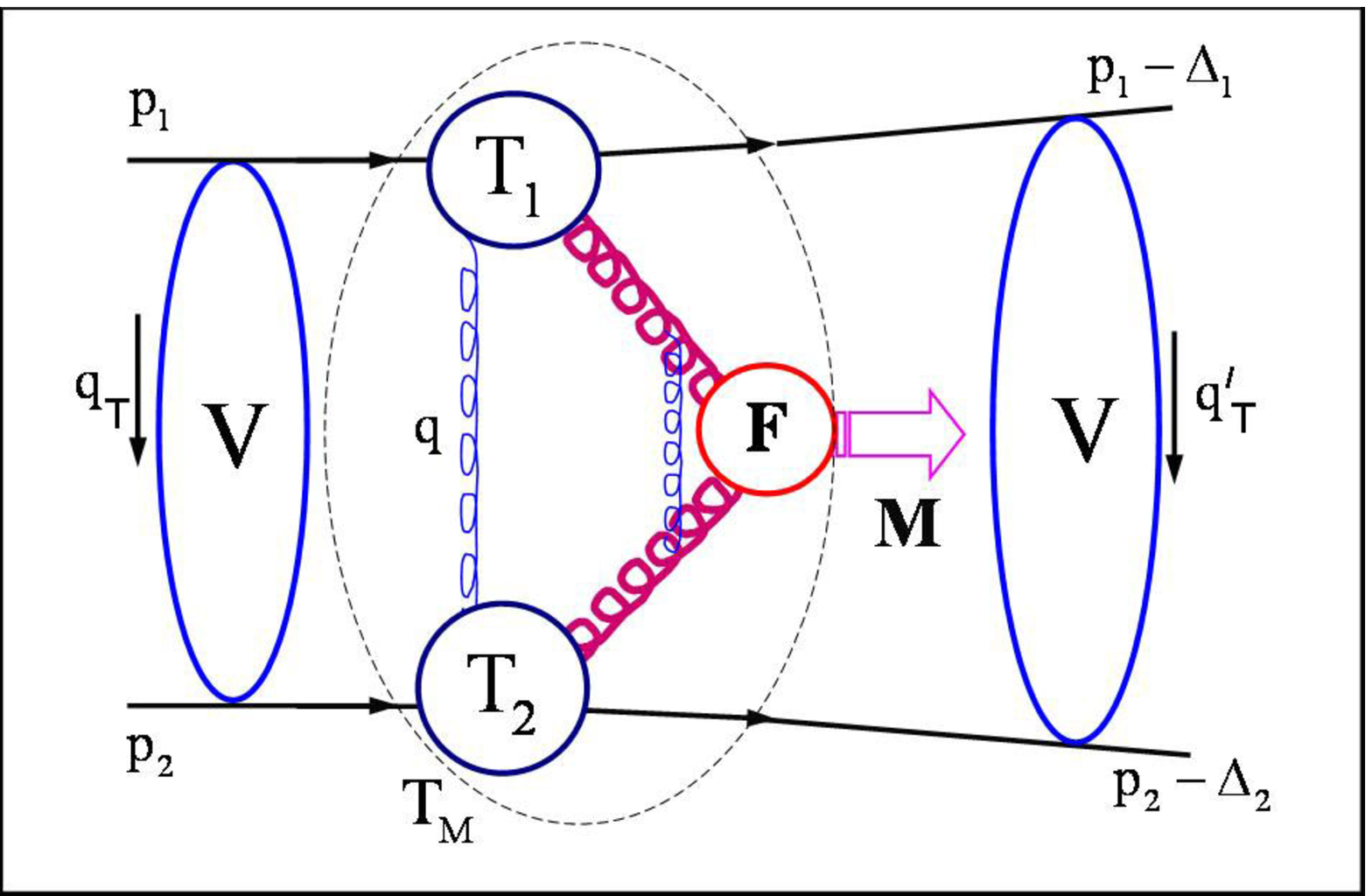}}}
\vskip -5.15cm
\hskip 8cm \mbox{b)}\vbox to 5cm {\hbox to 7cm{\epsfxsize=7cm\epsfysize=5cm\epsffile{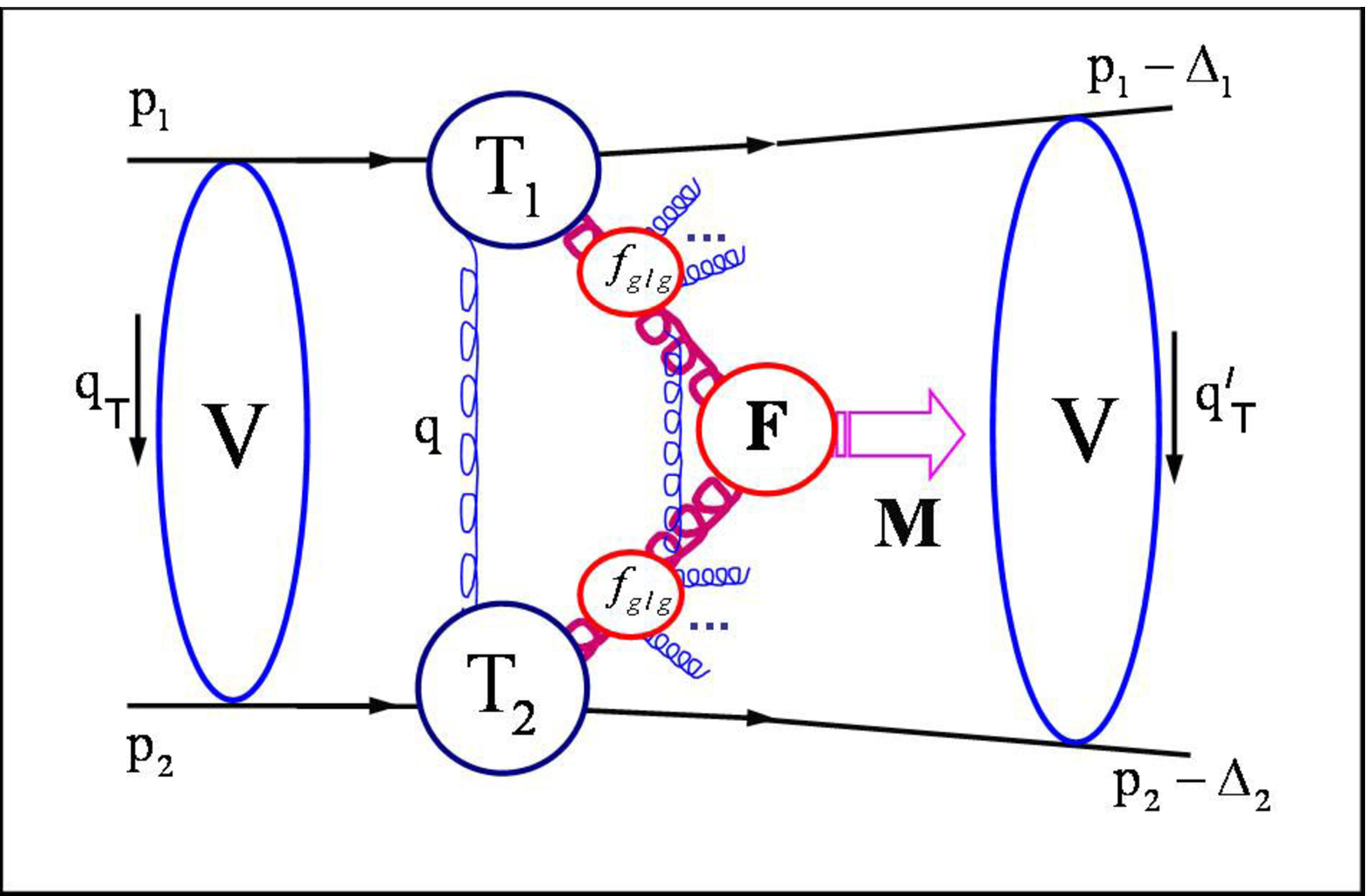}}}
\caption{a) The process $p+p\to p+M+p$;  b) The process $p+p\to p+\{ X M Y\}+p$. $X$ and $Y$ are "soft" gluons from $f_{g/g}$. }
\end{figure}

The exclusive double diffractive process is related to the
dominant amplitude of the exclusive two-gluon production. Driving
mechanism of this processes is the Pomeron. For calculation of cross-sections we use the method developed in Refs.~\cite{menu},\cite{2}. It is based on the extension of the Regge-eikonal approach, and succesfully used for the description of the HERA~\cite{3},\cite{4} and $p+p({\bar p})\to p+p({\bar p})$~\cite{5} data. 

In the framework of this approach, amplitudes for EDDE and SIDDE can be sketched
as shown in Fig.~1.

In the model we consider the kinematical region which is typical for diffraction:
\begin{equation}
\label{tlimits}
0.001\; GeV^2\le |t_{1,2}|\le 5\; GeV^2\;{,} 
\end{equation}
\begin{equation}
\label{xlimits}
\xi_{min}\simeq\frac{M^2}{s}\le \xi_{1,2}\le \xi_{max}=0.3\;,
\end{equation} 
where $t_{1,2}=\Delta^2_{1,2}$ are transfer momenta squared and $\xi_{1,2}=\Delta P_{L\; 1,2} /\sqrt{s}$, $\Delta P_{L\; 1,2}$ are longitudinal momentum transfers of protons.

The off-shell gluon-proton amplitudes $T_{1,2}$ are obtained in
the extended unitary approach~\cite{Petrov:95}.
After the tensor contraction of
these amplitudes with the gluon-gluon fusion vertex, the
full "bare" amplitude $T_M$ depicted in Fig.~1a inside the dashed oval
looks like
\begin{equation}\label{Tpp_pXp}
T_M = \frac{2}{\pi} \, c_{gp}^2 \, e^{b(t_1+t_2)}
\left(-\frac{s}{M^2}\right)^{\alpha_P(0)} F_{gg \to M} \, I_s \;.
\end{equation}
Here
\begin{eqnarray}\label{slope}
b &=& \alpha^{\prime}_P(0) \ln \left( \frac{\sqrt{s}}{M} \right)
+ b_0 \; ,
\\
b_0 &=& \frac{1}{4} \, (\frac{r^2_{pp}}{2} + r^2_{gp}) \; ,
\end{eqnarray}
with the parameters of the "hard" Pomeron trajectory, that
appears to be the most relevant in our case, presented in the
Table~\ref{tab:hpomeron}. The last factor in the r.h.s. of
~(\ref{Tpp_pXp}) is
\begin{eqnarray}\label{T12cor}
I_s &=& \int\limits_{0}^{\mu^2}\frac{dl^2}{l^4} \, F_s(l^2)
\left( \frac{l^2}{s_0 + l^2/2} \right)^{\alpha_P(t_1)+\alpha_P(t_2)}
(1+h(v,t_1))
(1+h(v,t_2))\; ,\\
\label{T12cor_h}
h(v,t)&=& \sum_{n=2}^{\infty} \frac{(-1)^{n-1}}{n!\cdot n}
\left(  \frac{c_{gp}}{8\pi\; b_1(v)}  
\exp\left[-\frac{i\pi(\alpha_P(0)-1)}{2} \right] v^{\alpha_P(0)-1} 
\right)^{n-1}\cdot\nonumber\\
&\cdot&\exp\left[ \frac{b_1(n-1)}{n}|t| \right]\; ,\\
v&=& \frac{\sqrt{s}}{M}\frac{l^2}{s_0 + l^2/2}\; ,\\
\label{T12cor_b1}
b_1&=& \alpha^{\prime}_P(0) \ln v +b_0\; .
\end{eqnarray}
Here $l^2=-q^2\simeq{\vec{\bf q}}^2$, $\mu=M/2$, and $s_0$ is a scale
parameter of the model which is also used in the global fitting of
the data on $pp$ ($p\bar{p}$) scattering for on-shell
amplitudes~\cite{5}. The fit gives $s_0 \simeq 1$~GeV$^2$. In~(\ref{T12cor}) we take into account rescattering corrections for "soft"
amplitudes $T_{1,2}$, which play role in the investigations of diffractive pattern~\cite{diffpatterns},~\cite{Note07022} for $t>1$~GeV and give very small contribution to the total cross-section. Below we perform also calculations of the so called "soft survival probability", which corresponds to "soft" rescattering corrections
in the initial and final states.

If we take into account the emission of virtual "soft" gluons,
while prohibiting the real ones, that could fill rapidity gaps, it
results in a Sudakov-like suppression~\cite{KMR3:sudakov}:
\begin{eqnarray}\label{sudakov}
F_s(l^2,\mu^2) &=& \exp\left[ -
\int\limits_{l^2}^{{\mu}^2} \frac{d p_T^2}{p_T^2} \,
\frac{\alpha_s({p_T}^2)}{2\pi} \int\limits_{\Delta}^{1-\Delta} z P_{gg}(z) d z
+\int\limits_0^1\sum_q P_{qg}(z)dz
\right] \; ,\\
P_{gg}(z)&=&6\frac{(1-z(1-z))^2}{z(1-z)}\; ,\\
\Delta&=&\frac{p_T}{p_T+\mu}\; .
\end{eqnarray}

The "hard" part
of the EDDE amplitude, $F_{gg\to M}$, is the usual gluon-gluon
fusion amplitude calculated perturbatively in the SM or in its
extensions.

\begin{table}[h!]
\begin{center}
\caption{Phenomenological parameters of the "hard" Pomeron
trajectory obtained from the fitting of the HERA and Tevatron data
(see~\cite{menu}, \cite{4},\cite{Note07022}), and data on $pp$
($p\bar{p}$) scattering~\cite{5}.The value of the $c_{gp}$ is corrected
in accordance with the latest data from CDF~\cite{CDF2006}.}
\bigskip \bigskip
  \begin{tabular}{||c|c|c|c|c||}
  \hline
  $\alpha_P(0)$ & $\alpha_P^{\prime}(0)$, GeV$^{-2}$ &  $r^2_{pp}$, GeV$^{-2}$
  & $r^2_{gp}$, GeV$^{-2}$ & $c_{gp}$
  \\ \hline
  1.203 &  0.094 &  2.477 &  2.54 & 3.2$\pm$0.5
  \\ \hline
  \end{tabular}
\label{tab:hpomeron}
\end{center}
\end{table}

The data on total cross-sections demand unambiguously the Pomeron
with larger-than-one intercept, thereof the need in unitarization, i.e.
taking into account "soft" rescattering in the initial and final states. The 
full amplitude in the Fig.~1a with 
unitary corrections, $T^{unit}_M$, is given by the following analytical
expressions:
\begin{eqnarray}\label{ucorr}
T^{unitar}_M(p_1, p_2, \Delta_1, \Delta_2) &=& \frac{1}{16\,s
s^{\prime}} \int \frac{d^2\vec{\bf q}_T}{(2\pi)^2} \,
\frac{d^2\vec{\bf q}^{\prime}_T}{(2\pi)^2} \; V(s, \vec{\bf
q}_T) \;
\nonumber \\
&\times& T_M( p_1-q_T, p_2+q_T,\Delta_{1T}, \Delta_{2T}) \,
V(s^{\prime}, \vec{\bf q}^{\prime}_T) \;,
\\
V(s, \vec{\bf q}_T) &=& 4s \, (2\pi)^2 \, \delta^2(\vec{\bf
q}_T) + 4s \!\! \int d^2\vec{\bf b} \, e^{i\vec{\bf q}_T
\vec{\bf b}} \left[e^{i\delta_{pp\to pp}}-1\right]\;,
\end{eqnarray}
where $\Delta_{1T} = \Delta_{1} -q_T - q^{\prime}_T$, $\Delta_{2T}
= \Delta_{2} + q_T + q^{\prime}_T$, and the eikonal function
$\delta_{pp\to pp}$ can be found in Ref.~\cite{5}. Left
and right parts of the diagram in Fig.~1a denoted by
$V$ represent these "soft" re-scattering effects. As was shown in~
\cite{EDDE:glueballs}, these "outer" unitary corrections
strongly reduce the value of the corresponding cross-section and
change an azimuthal angle dependence.

To calculate differential and total cross-sections for exclusive processes
we can use the formula
\begin{eqnarray}
M^2\frac{d\sigma^{EDDE}}{dM^2\; dy\; d\Phi_{gg\to M}}|_{y=0}&=&\hat{L}^{EDDE}
\frac{d\hat{\sigma}^{J_z=0}}{d\Phi_{gg\to M}}\; ,\\
\label{lumEDDE}
\hat{L}^{EDDE}&=& \frac{c_{gp}^4}{2^5\pi^6}
\left( \frac{s}{M^2} \right)^{2(\alpha_P(0)-1)} \frac{1}{4b^2} I_s
S^2\; ,\\
\label{softsurv}
S^2 &=& \frac{\int d^2\vec{{\bf\Delta}}_1d^2\vec{{\bf\Delta}}_2 |T^{unitar}_M|^2}{\int d^2\vec{{\bf\Delta}}_1d^2\vec{{\bf\Delta}}_2 |T_M|^2}\; ,
\end{eqnarray}
where $d\hat{\sigma}^{J_z=0}/d\Phi_{gg\to M}$ is the "hard" 
exclusive singlet gluon-gluon fusion cross-section and $S^2$ is the so called "soft" 
survival probability. For fixed energy $S^2$ is approximately constant. 

We can extend our approach to the case of additional "soft" radiation
in the central rapidity region. This process is depicted 
in~Fig.~1b. First of all we have to calculate
unintegrated gluon distribution $\hat{f}_{g/g}(x,k^2_t,\mu^2)$ inside a gluon. For this task we use the method similar to the one presented in~\cite{weber} (convenient for
further Monte-Carlo simulation) and~\cite{KMRunintegr}. Generation is
performed step by step by the initial gluon splitting. In this case Sudakov-like
form factor $F_s$ in the above formulaes for cross-sections has to be replaced by the new one~\cite{Note07022}
\begin{equation}
\label{sudSIDDE}
F^{SI}_s= \left[ \int\limits^{k_{T,max}^2} \frac{dk^2_t}{k^2_t} \int dx\; 
x^{2(\alpha_P(0)-1)}
\hat{f}_{g/g}(x,k_{t}^2,\mu^2) \right]^2\;,
\end{equation}
and partonic exclusive $gg\to M$ cross-section has to be replaced by the inclusive one.
Since there
are some theoretical uncertainties concerning the choice of the so called "factorization"
scale $\mu$, this question has to be considered explicitly by methods of
the renormalization theory. In the next version of the generator we will present more exact calculations of $\hat{f}_{g/g}(x,k_{t}^2,\mu^2)$ based on the Bethe-Salpeter equation, taking into account also
color correlations, which are included in this version in a more simple way. 

\section{Main subroutines and common blocks}

{\bf Basic functions and subroutines}:

\begin{itemize}

\item Subroutine {\tt EDDETTPHI(NX,MX,GT1,GT2,GFI0)} generate
$t_1$, $t_2$, $\phi$ - distribution for different masses {\tt MX} of the
central system and different quantum numbers of the system:
\begin{itemize}
\item[{\tt NX}$=1$] $\longrightarrow 0^{++}$ (default)
\item[{\tt NX}$=2$] $\longrightarrow J_z=0,\;\pm 2$
\item[{\tt NX}$=3$] $\longrightarrow 0^{-+}$ 
\item[{\tt NX}$=4$] $\longrightarrow $ glueball
\item[{\tt NX}$=5$] $\longrightarrow J_z=\pm 2$
\end{itemize}

\item Function {\tt EDDEX(MX)} generate $\xi$ distribution for a final proton.

\item Functions of the type {\tt DCSxx(N,M,ETA)} 
or {\tt DCSQQ(MQ,N,M,ETA)} compute exclusive differential
partonic cross section 
$$
\frac{d\hat{\sigma}^{J_z=0}}{d\eta^*}\;,
$$
where $\eta^*$ is the pseudorapidity of the final parton
in the initial gg central mass frame. {\tt MQ} is the mass of a 
quark. Functions {\tt CSQQ(MQ,M)} 
or  {\tt CSxx(M)} are integrated exclusive gluon-gluon fusion cross-sections. 
{\tt xx}$=${\tt GG},{\tt 2GAM},{\tt 3G},{\tt QQG} denotes the final
system ($gg$,$\gamma\gamma$,$ggg$,$Q\bar{Q}g$). {\tt N} is an auxiliary integer
parameter. If {\tt N}$=0$ then we obtain exact cross-section, else functions
return an upper estimations for cross-sections.

\item Functions of the type {\tt DCSxxSI(N,M,ETA)} 
or {\tt DCSQQSI(MQ,N,M,ETA)} are inclusive differential
partonic cross-section (without the rule $J_z=0$). Functions {\tt CSQQSI(MQ,M)} 
or  {\tt CSxxSI(M)} are integrated inclusive gluon-gluon fusion cross-sections.

\item Subroutines of the type {\tt GENERyyxx(GMGG,GETAJ)}, {\tt GENERyyQQ(MQ,GMGG,GETAJ)}, \linebreak
{\tt GENEREX3G(GMGG,GETAJ,GX3,GPT3,GFI3)} ($ggg$ distributions),\\
{\tt GENEREXQQG(GMGG,GETAJ,GX3,GFIS,GTHETAS)} ($Q\bar{Q}g$ distributions) 
generate distributions in $\eta^*$({\tt GETAJ}) and mass of the central system ({\tt GMGG}). Also
{\tt GX3}, {\tt GPT3}, {\tt GFI3}, {\tt GFIS}, {\tt GTHETAS} are variables for the 3rd jet in the 3-jet kinematics. {\tt yy}$=${\tt EX}, {\tt SI} means Exclusive and Semi-inclusive events. {\tt xx}$=${\tt GG}, {\tt 2GAM} ($gg$ and $\gamma\gamma$ final systems).

\item Subroutine {\tt SICASCAD2(MC,N1,PG1,P1,N2,PG2,P2,MX,NFAIL)}
generates two correlated "soft" gluonic systems X and Y in the process $p+p\to p+\{X M Y\}+p$. {\tt MC} is the mass of the final system (Higgs boson,Radion,graviton,glueball,$\chi_{c,b}$,$gg$,$Q\bar{Q}$,$\gamma\gamma$), {\tt MX} is the mass of the system $\{ X M Y\}$, {\tt N1}, {\tt N2} are numbers of gluons
in $X$ and $Y$. {\tt PG1(5,500)}, {\tt PG2(5,500)} are four momenta of gluons in $X$ and $Y$, and {\tt P1(5)}, {\tt P2(5)} are momenta of "hard" gluons in the process $gg\to M$.

\item Function {\tt SOFTSURV(N,MGG)} returns the "soft survival probability" factor
$S^2$. {\tt MGG} is the central mass.

\item Function {\tt M2DLUMDM2(N1,N2,MGG)} returns "luminosity" $\hat{L}^{EDDE,\;SIDDE}$ integrated in the rapidity of the central system. {\tt N1}$=0$ for EDDE, {\tt N1}$=1$ for SIDDE. {\tt N2}$=0$ for resonances, {\tt N2}$=1$ for jets and $\gamma\gamma$.

\item Subroutine {\tt EDDERS1C(NRS,RSXI,RSGAM,RSMH,RSMR,RSMOBS,RSWD,BR)} returns mas\-ses, 
widths and branching fractions for resonances of the RS1 model~\cite{RS1}.Parameter
{\tt NRS} switches between two mass states $H^*$ (Higgs boson, {\tt NRS}$=1$) and $R^*$ (Radion, {\tt NRS}$=2$). {\tt RSXI}, {\tt RSGAM}, {\tt RSMH}, {\tt RSMR} are
input parameters for the RS1 model (mixing parameter $\xi$, scale parameter $\gamma$, "bare" masses $m_h$, $m_r$ before mixing)~\cite{RS1}. {\tt RSMOBS}, {\tt RSWD}, {\tt BR} are the mass, width and branching fraction into $b\bar{b}$ state
for $H^*$ or $R^*$.

\item Function {\tt EDDECS(IP)} retuns total cross-section of the corresponding process:
\begin{itemize}
\item[{\tt IP}$=400$:] EDDE, resonance production (Standard Model Higgs boson, $H^*$ and $R^*$ of RS1 model, $\chi_{c,b}$, glueball, graviton). Partonic $gg\to H$ cross-section could be found in~\cite{H}. Other EDDE cross-sections were calculated in~\cite{menu},\cite{graviton}.
\item[{\tt IP}$=401$:] EDDE, $Q\bar{Q}$ production.
\item[{\tt IP}$=402$:] EDDE, $gg$ production.
\item[{\tt IP}$=403$:] EDDE, $\gamma\gamma$ production.
\item[{\tt IP}$=404$:] EDDE, $Q\bar{Q}g$ production.
\item[{\tt IP}$=405$:] EDDE, $ggg$ production.
\item[{\tt IP}$=406$:] SIDDE, resonance production.
\item[{\tt IP}$=407$:] SIDDE, $Q\bar{Q}$ production.
\item[{\tt IP}$=408$:] SIDDE, $gg$ production.
\item[{\tt IP}$=409$:] SIDDE, $\gamma\gamma$ production.
\end{itemize}

\item Subroutine {\tt EDDEEVE} returns an event in PYTHIA~\cite{pythia} format
according to the value of the process identification variable (see data card below).

\item Subroutine {\tt EDDEINI} makes initialization of the generator and provide the
interface with PYTHIA.

\item Subroutine {\tt EDDEPUTDAT} fills all the internal common blocks. It is called
from {\tt EDDEINI}.

\end{itemize}

All the cross-sections are calculated in mb.

\phantom{xxx}

{\bf EDDE common blocks}:

\begin{itemize}

\item Fundamental constants:\\
\\
{\tt \phantom{X}COMMON/EDDEFUND/ MNI,REI,PI,CSMB,LAMQCD,TF,CF,BF0,BF1,NF,NC,NLOSW}\\
{\tt \phantom{X}...\\
      \phantom{X}DATA PI/3.141592654D0/, CSMB/0.38D+00/ \\
      \phantom{X}DATA REI/(1.D0,0.D0)/,MNI/(0.D0,1.D0)/ \\
      \phantom{X}DATA NC/3/,NF/5/,NLOSW/1/,LAMQCD/0.5D-01/ \\
      \phantom{X}DATA TF/0.5D0/ \\
}
\\
{\tt TF,CF,BF0,BF1,NF,NC} are SU(3) constants\\
{\tt NLOSW} is a switch parameter. For zero value $\alpha_s$ is calculated in the
Leading Order, for unit value we have NLO calculations.

\item Parameters for the Regge-eikonal model and Regge kinematics:\\
\\
{\tt \phantom{X}COMMON/EDDESOFT/ CP(3),DP(3),RP(3),RG(3),AP(3),\\
    \&   T1MIN,T1MAX,T2MIN,T2MAX,FKK,CGP,NAPR,NFI }\\ 
\\    
Parameters of trajectories: couplings {\tt CP}, intercepts {\tt DP}$=\alpha_{IP}(0)-1$, 
{\tt RP}$=r_{pp}^2$, {\tt RG}$=r_{gp}^2$, slopes {\tt AP}$=\alpha_{IP}^{\prime}(0)$\\
\\
{\tt \phantom{X}... \\
      \phantom{X}DATA CP/0.5300D+02,0.9700D+01,0.1670D+01/ \\
      \phantom{X}DATA DP/0.5800D-01,0.1670D+00,0.2030D+00/ \\
      \phantom{X}DATA RP/0.6300D+01,0.3100D+01,0.2480D+01/ \\
      \phantom{X}DATA RG/0.6300D+01,0.3100D+01,0.2540D+01/ \\
      \phantom{X}DATA AP/0.5600D+00,0.2730D+00,0.9400D-01/ \\
}
\\
$t_{1,2}$ limits:\\
\\
{\tt                 
      \phantom{X}DATA T1MIN/0.1D-02/, T1MAX/0.7D+01/ \\
      \phantom{X}DATA T2MIN/0.1D-02/, T2MAX/0.7D+01/ \\
}
\\
Number of terms in the expansion:\\
\\
{\tt 
      \phantom{X}DATA NAPR/9/ \\
}
\\
Main constant of $gp\to gp$ amplitude:\\
\\
{\tt  \phantom{X}DATA CGP/0.316D+01/ ! (3.2+-0.5)  \\
}
\\
{\tt NFI}$=${\tt IPAR 5} from the data card, it is the code for azymuthal distribution. 

\item Kinematical limits for $\xi_{1,2}$:\\
\\
{\tt \phantom{X}COMMON/EDDETOT/ XI1MIN,XI2MIN,XI1MAX,XI2MAX}\\
{\tt  \phantom{X}...  \\
      \phantom{X}DATA XI1MIN/0.1D-04/,XI2MIN/0.1D-04/ \\
      \phantom{X}DATA XI1MAX/0.1D0/,XI2MAX/0.1D0/ \\
}      

\item Parameters for "hard" gluon-gluon fusion cross-sections:\\
\\
{\tt \phantom{X}COMMON/EDDEHARD/ MGGCUT,ETJCUT,MXMAX,\\
     \& ETAJMAX,PLUM,PSURV,PSUD,ETASIMAX,SQS,\\
     \& PSIDD1,PSIDD2\\
}\\     
{\tt MGGCUT}$=2*${\tt ETJCUT}, {\tt ETJCUT}$=${\tt RPAR 2} is the cut on $m_T=\sqrt{E_T^2+m_{part}^2}$ for final "hard" parton. For gluons and photons $m_part=0$, i.e. {\tt ETJCUT} is the cut on the transverse momenta.\\
{\tt MXMAX}$=${\tt 500.D0} GeV is the upper bound for the mass of the central system  in the generator.\\
{\tt ETAJMAX}$=\max|\eta^*_{part}|$ is the upper bound for the pseudorapidity of final "hard" partons in the central mass frame of initial gluons.\\
{\tt ETASIMAX}$=${\tt RPAR 9} is the pseudorapidity interval for "soft" radiation.\\
{\tt SQS}$=\sqrt{s}=${\tt RPAR 1}.\\
Other parameters are used for different auxiliary distributions.\\

\item Auxiliary parameters for 3-jet kinematics:\\
\\
{\tt \phantom{X}COMMON/EDDE3JP/ DER3J,XMAX3J,PAR3G(5)}\\

\item Parameters for RS1 model~\cite{RS1}:\\
\\
{\tt \phantom{X}COMMON/EDDERS1/ RSXI0,RSGAM0,RSMH0,RSMR0,NRS0}\\
\\
{\tt RSXI0}$=${\tt RPAR 5}$=\xi$ is the mixing parameter,\\
{\tt RSGAM0}$=${\tt RPAR 6}$=\gamma=246$~GeV$/\Lambda_{\phi}$ is the scale parameter,\\
{\tt RSMH0}$=${\tt RPAR 7}$=m_h$ and {\tt RSMR0}$=${\tt RPAR 8}$=m_r$ are "bare" masses of the model,\\
{\tt NRS0}$=${\tt IPAR 6} is the switch parameter between SM Higgs boson ({\tt IPAR 6}$=0$), $H^*$ ({\tt IPAR 6}$=1$), $R^*$ ({\tt IPAR 6}$=2$).

\item Additional global parameters:\\
\\
{\tt \phantom{X}COMMON/EDDEOTHER/ KCP,IPROC,AM0,AMP,S,MQ}\\
\\
{\tt KCP}$=${\tt IPAR 4} is the code of the central particle in the process number 400 and 406,\\
{\tt IPROC}$=${\tt IPAR 1} is the number of process (see above the parameter {\tt IP} of the function {\tt EDDECS}),\\
{\tt AM0}$=${\tt RPAR 3} is the mass of the central particle,\\
{\tt AMP} is the proton mass, {\tt S}$=${\tt SQS}$*${\tt SQS},\\
{\tt MQ}$=${\tt RPAR 10} is the mass of heavy quark (b-quark by default).\\
\\
\item Data tables for luminocities and distribution functions:\\
\\
{\tt \phantom{X}COMMON/EDDETAB1/ LUM1(480),FLUM1(30,16),\\
     \& DX1,DY1,X01,Y01\\
     \phantom{X}...\\
    \phantom{X}COMMON/EDDETAB2/ RDI3G(630),FRDI3G(30,21),\\
     \& DX2,DY2,X02,Y02\\
     \phantom{X}...\\
      \phantom{X}/EDDETAB3/ RI3GA(480),RI3GB(480),\\
     \& FRI3GA(30,16),FRI3GB(30,16),DX3,DY3,X03,Y03\\     
\phantom{X}...\\
       \phantom{X}COMMON/EDDETAB4/ FT4(10201),FMT4(101,101),\\
     \& DX4,DY4,X04,Y04 \\
\phantom{X}...\\
       \phantom{X}COMMON/EDDETAB5/ DFT5(10201),DFMT5(101,101),\\
     \& DX5,DY5,X05,Y05\\ }  
\end{itemize}

\section{Data card}

User can manage the process of generation with the help of the data card. Parameters
that could be changed and their default values:

\begin{itemize}
\item {\tt IPAR 1=403}: identification number of the process ({\tt IP} in the function {\tt EDDECS}). 
\item {\tt IPAR 2=10000}: number of generated events. 
\item {\tt IPAR 4=25}: code of the central particle for processes number 400 and 406 (according to PYTHIA codes for particles, for example, 25 for Higgs boson and RS1 particles, 50 for glueball etc).
\item {\tt IPAR 5=1}: parameter to switch between different azymuthal distributions ({\tt NX} in the subroutine {\tt EDDETTPHI}).
\item {\tt IPAR 6=0}: number of particle for the process 400 or 406 and {\tt IPAR 4=25}. {\tt IPAR 6=0}: Standard Model Higgs boson; {\tt IPAR 6=1}: $H^*$ ("Higgs boson") mass state of the RS1 model; {\tt IPAR 6=2}: $R^*$ (Radion) mass state of the RS1 model. 
\item {\tt RPAR 1=14000.0E+00}: (in GeV) center mass energy $\sqrt{s}$.
\item {\tt RPAR 2=25.0E+00}: (in GeV) transverse mass cut $m_T=\sqrt{E_T^2+m_{part}^2}$ for the "hard" final parton of the central system. For gluons and photons this parameter is equal to $E_T$ cut.
\item {\tt RPAR 3=120.0E+00}: (in GeV) mass of the central particle for processes number 400 and 406.
\item {\tt RPAR 5=0.16E+00}: mixing parameter $\xi$ of the RS1 model, typical values $-0.2\to 0.2$.
\item {\tt RPAR 6=0.246E+00}: scale parameter $\gamma=246$~GeV$/\Lambda_{\phi}$, where $\Lambda_{\phi}$ has typical values $1000\to 5000$~GeV. 
\item {\tt RPAR 7=150.0E+00}: (in GeV) "bare" mass of Higgs boson before mixing $m_h$. Typical values $120\to 180$~GeV.
\item {\tt RPAR 8=110.0E+00}: (in GeV) "bare" mass of Radion before mixing $m_r$. Typical values $\sim 100\to 300$~GeV.
\item {\tt RPAR 9=5.0E+00}: pseudorapidity interval for "soft" radiation $X$ and $Y$ in the SIDDE, $-${\tt RPAR 9}$/2$$<\eta<${\tt RPAR 9}$/2$.
\item {\tt RPAR 10=4.8E+00}: (in GeV) mass of the final "hard" quark in the central system.
\end{itemize}

Some PYTHIA definitions:\\
\begin{verbatim}
PYTHIA
C
MRPY 1=77123456          ! State of random number generator   
MSEL 0                 !full user control
C...some PYTHIA definitions...
MSTP 61 =1             ! Initial-state QCD and QED radiation
MSTP 71 =1             ! Final-state QCD and QED radiation
MSTP 81 =1             ! multiple interaction
MSTP 111=1             ! fragmentation and decay
MSTP 122=0             ! switch off X section print out
C...Higgs decay definition...
MDME 210,1 =0           ! h0 -> d dbar
MDME 211,1 =0           ! h0 -> u ubar
MDME 212,1 =0           ! h0 -> s sbar
MDME 213,1 =0           ! h0 -> c cbar
MDME 214,1 =1           ! h0 -> b bbar 
MDME 215,1 =0           ! h0 -> t tbar  
MDME 216,1 =-1          ! h0 -> b' b'bar
MDME 217,1 =-1          ! h0 -> t' t'bar
MDME 218,1 =0           ! h0 -> e+e-
MDME 219,1 =0           ! h0 -> mu+mu- 
MDME 220,1 =0           ! h0 -> tau+tau-
MDME 221,1 =-1          ! h0 -> tau'+ tau'-
MDME 222,1 =0           ! h0 ->  gg 
MDME 223,1 =0           ! h0-> gamma gamma
MDME 224,1 =0           ! h0 -> gamma Z0  
MDME 225,1 =0           ! h0 -> Z0 Z0  
MDME 226,1 =0           ! h0 -> W+W-  
\end{verbatim}


\section{Results from EDDEv2.1}

\begin{figure}[h]
\label{samples}
\mbox{a)}\hskip -0.5cm
\vbox to 7cm {\hbox to 7cm{\epsfxsize=7cm\epsfysize=7cm\epsffile{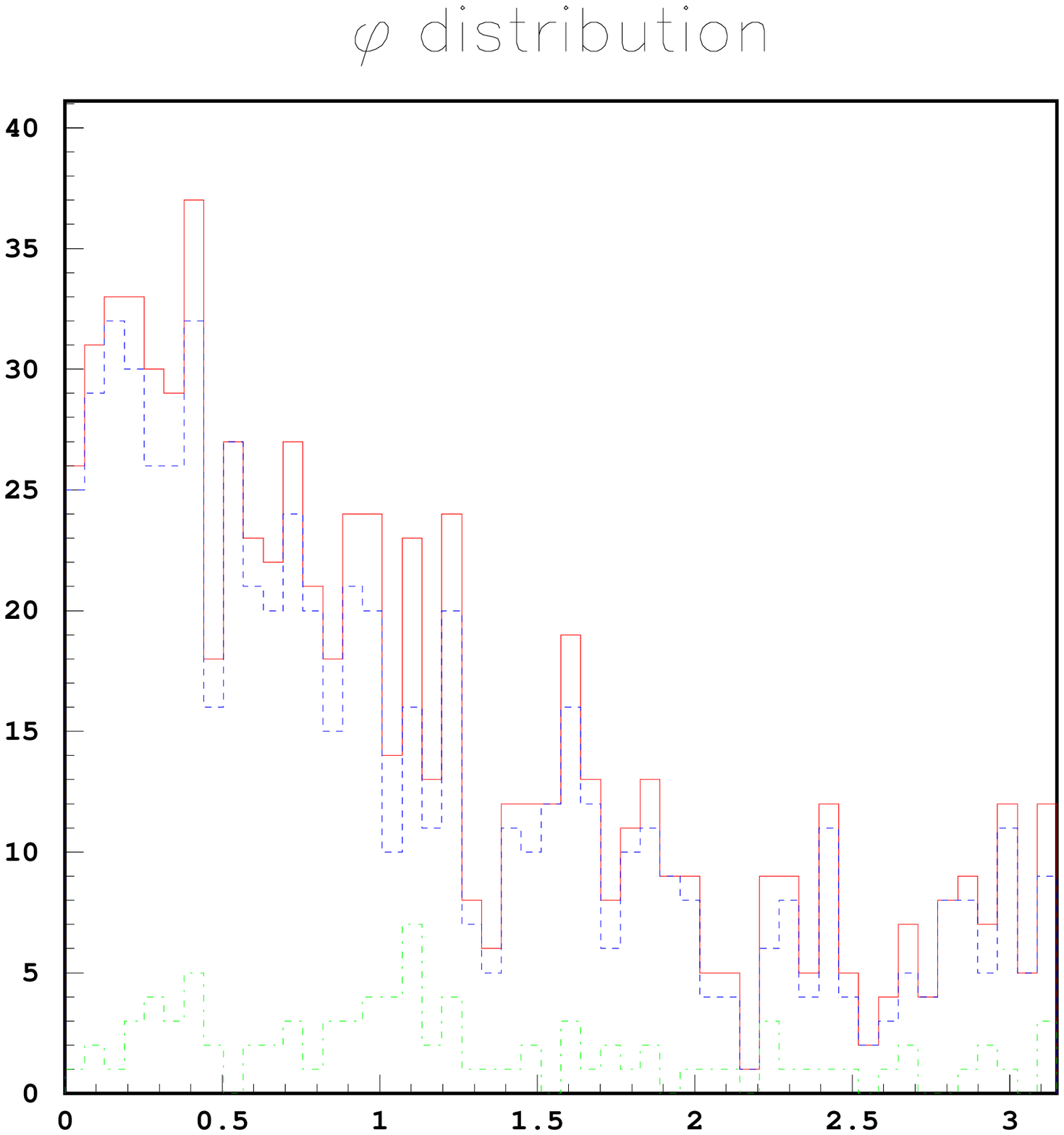}}}
\mbox{b)}
\vskip -7cm
\hskip 7cm
\vbox to 7cm {\hbox to 7cm{\epsfxsize=7cm\epsfysize=7cm\epsffile{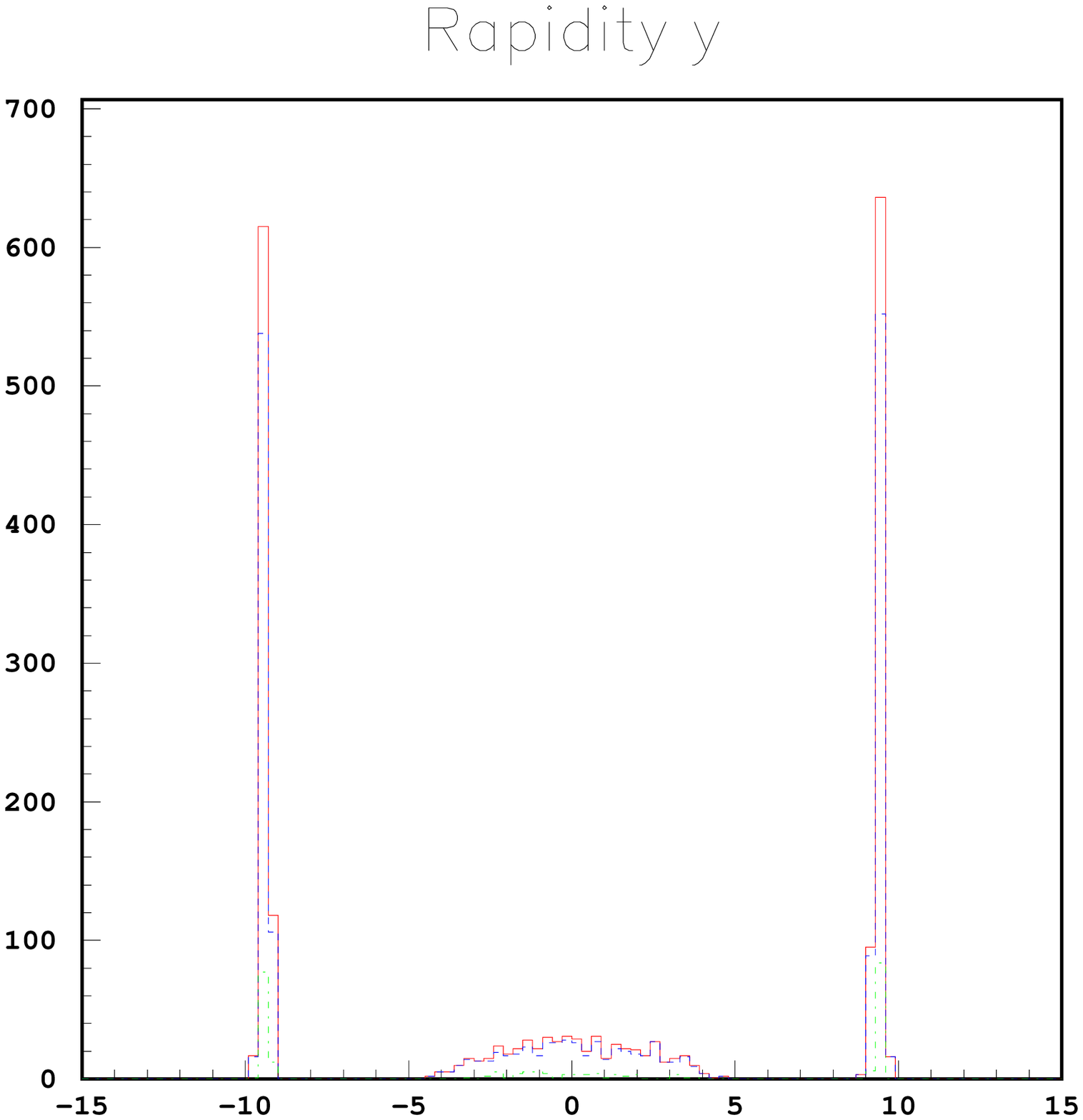}}}
\\
\mbox{c)}\hskip -0.5cm
\vbox to 7cm {\hbox to 7cm{\epsfxsize=7cm\epsfysize=7cm\epsffile{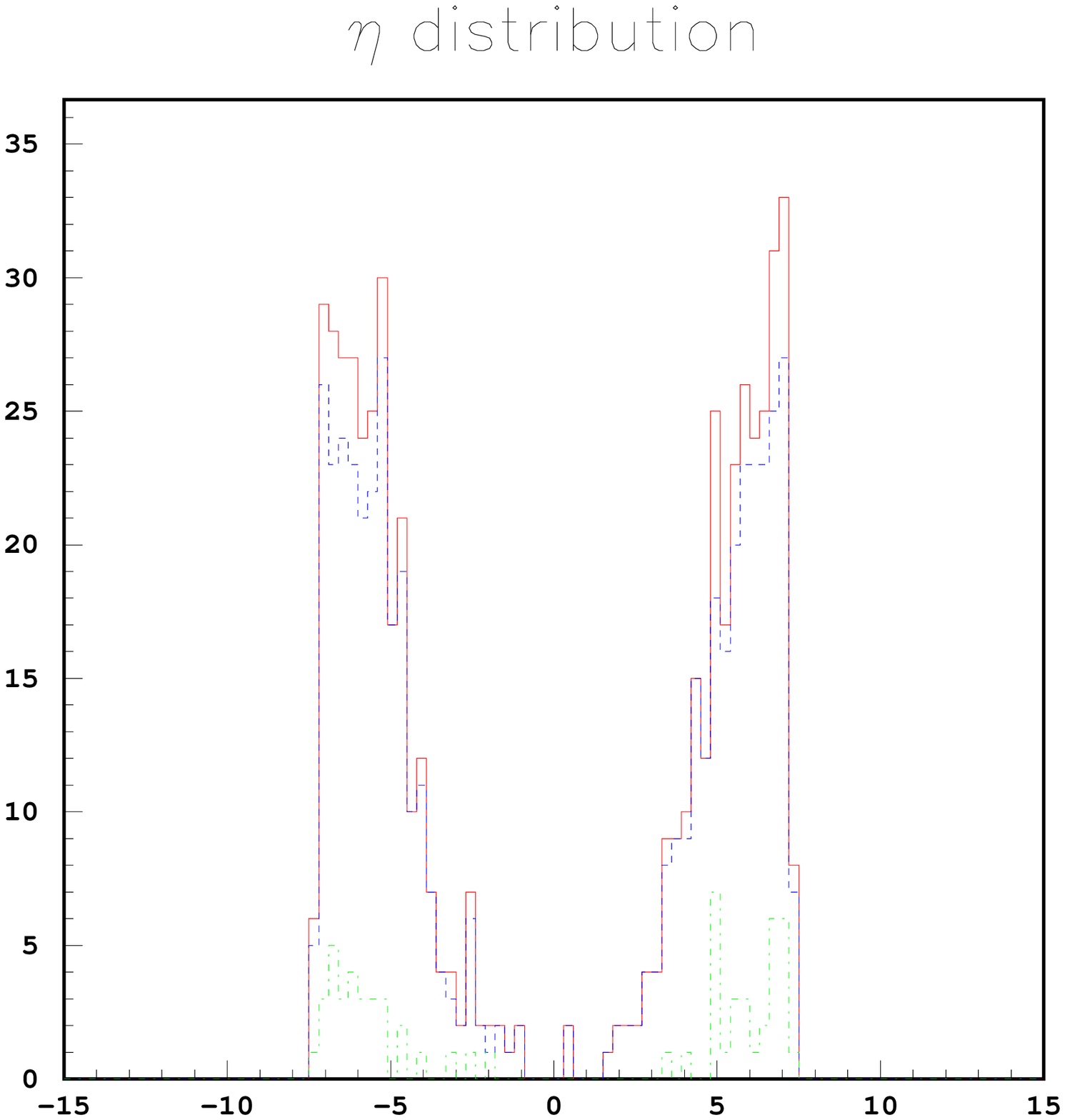}}}
\vskip -7cm
\hskip 7cm
\mbox{d)}
\vbox to 7cm {\hbox to 7cm{\epsfxsize=7cm\epsfysize=7cm\epsffile{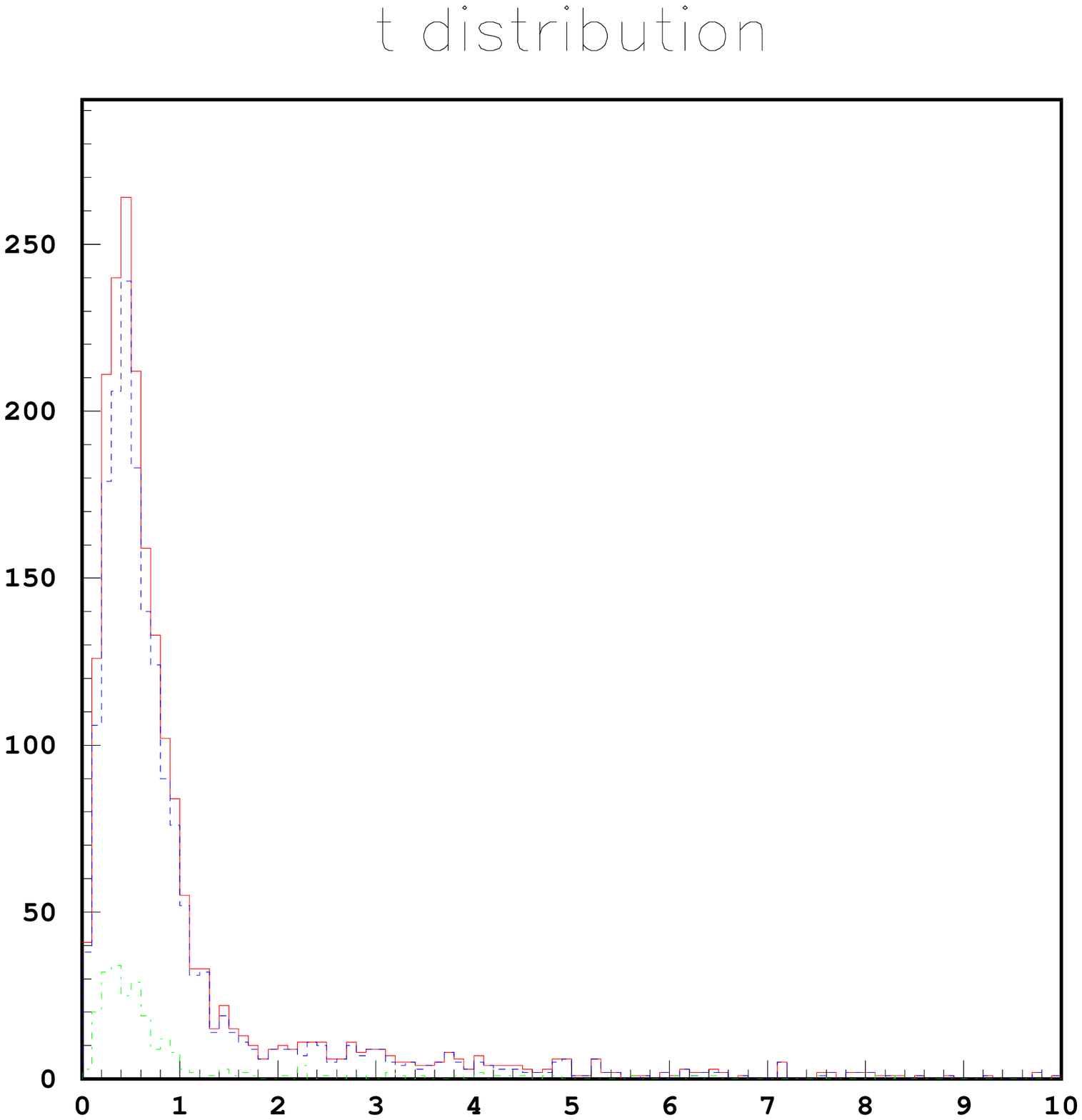}}}
\caption{Some distributions from EDDE. (90 signal and 660 backround events. $MTcut=25$~GeV). Solid curve represent signal+background, dashed one is the background, and dash-dotted is the signal.  a) azimuthal angle distribution for Higgs boson production; b) rapidity and c) pseudorapidity distributions; d) integrated t-distributions.}
\end{figure}

Here some samples from the work of the generator with PYTHIA~\cite{pythia} are presented (see Figs.~2a)-d).) 

The updated version of the generator EDDEv2.1 will be available on the web-page:
$$
http://sirius.ihep.su/cms/higgsdiff/diff.html
$$

\section*{Aknowledgements}

 This work is supported by grants INTAS-05-112-5481 and RFBR-06-02-16031.

\noindent Thanks to Kiril Datsko for the interface to the latest version of {\tt CMSSW}.
 


\begin{thebibliography}{9}

  \bibitem{diffpatterns} V.A. Petrov and R.A. Ryutin, CMS Internal Note 2006/051
  \bibitem{TDRS} M.Albrow et al.,CERN-LHCC-2006-039, CERN-LHCC-G-124, CERN-CMS-NOTE-2007-002;\\ 
  K. Eggert, M. Oriunno and M. Bozzo, \emph{TOTEM
  Technical Design Report}, CERN-LHCC-2004-002;\\
  M. Deile, Talk given at the Workshop ``Physics at LHC'' (July
2004, Vienna), arXive:hep-ex/0503042.
 \bibitem{menu} V.A. Petrov and R.A. Ryutin, JHEP {\bf 0408} (2004) 013.
  \bibitem {2} V.A. Petrov and R.A. Ryutin,  Eur. Phys. Journ. C {\bf 36} (2004) 509.  
  \bibitem {3} V.A. Petrov and A.V. Prokudin, Phys. Atom. Nucl. {\bf 62} (1999) 1562.
  \bibitem {4} V.A. Petrov, A.V. Prokudin and R.A. Ryutin, Czech. J. Phys. {\bf
55} (2005) 17.
  \bibitem {5} V.A. Petrov and A. V. Prokudin, Eur. Phys. J. C {\bf 23} (2002) 135.
  \bibitem{Petrov:95}
  V.A. Petrov, Talk given at the 6th Workshop on Elastic and
  Diffractive Scattering (20-24 Jun 1995, Blois, France). In
\emph{Blois 1995, Frontiers in strong interactions}, p. 139-143.
  \bibitem{Note07022} V.A. Petrov and R.A. Ryutin, CMS Internal Note 2007/022  
 \bibitem{KMR3:sudakov} V.A. Khoze, A.D. Martin and M.G. Ryskin, Eur. Phys. J. C 
{\bf 14} (2000) 525; \emph{ibid.} C {\bf 21} (2001) 99.
\bibitem{CDF2006} CDF Coll.: C. Mesropian, FERMILAB-CONF-06-464-E;\\
K. Goulianos, FERMILAB-CONF-06-464-E.
\bibitem{EDDE:glueballs}
V.A. Petrov, R.A. Ryutin,  A.E. Sobol and J.-P. Guillaud, JHEP {\bf 0506} (2005) 007. 
\bibitem{weber} G.Marchesini and B.R. Webber, Nucl.Phys. B {\bf 310} (1988) 461. 
\bibitem{KMRunintegr} M.A. Kimber, A.D. Martin, M.G. Ryskin, Phys.Rev. D 
{\bf 63} (2001) 114027. 
  \bibitem {RS1} L. Randall and R. Sundrum, Phys. Rev. Lett. {\bf 83} (1999) 3370;\\
  M. Chaichian, A. Datta, K. Huitu and Zenghui Yu, Phys.Lett. B {\bf 524} (2002) 161.
\bibitem {H} A. Kniehl, Phys. Rept. {\bf 240} (1994) 211.  
\bibitem{graviton} A.V. Kisselev, V.A. Petrov, R.A. Ryutin, Phys.Lett. B {\bf 630} (2005)  
100. 
\bibitem{pythia} T. Sjostrand et al., {\it PYTHIA}, Comp. Phys. Commun. {\bf 135} (2001) 238, hep-ph/0108264; hep-ph/0308153; JHEP {\bf 0605} (2006) 026 (recent version 6.4).
 
\end{thebibliography}
\end{document}